# A Tribute to Phil Bourne—Scientist and Human

Cameron Mura[1,*], Emma Candelier[1] and Lei Xie[2]

[1] School of Data Science; University of Virginia; Charlottesville, VA, 22903, USA
[2] Department of Computer Science; Hunter College, The City University of New York; New York, NY, 10065, USA
[*] Correspondence: cmura@virginia.edu

ABSTRACT: This Special Issue of *Biomolecules*, commissioned in honor of Dr. Philip E. Bourne, focuses on a new field of *biomolecular data science*. In this brief retrospective, we consider the arc of Bourne's 40-year scientific and professional career, particularly as it relates to the origins of this new field.

Phil, as he is known to all—from students to university presidents and beyond—is the founding Dean of the School of Data Science (SDS) at the University of Virginia (UVA). He previously served as the first Associate Director for Data Science at the U.S. National Institutes of Health (NIH), where he led a novel *Big Data to Knowledge* initiative [1]. Prior to the NIH, Phil had a highly productive and impactful 20-year career at the University of California, San Diego (UCSD), with close ties to the San Diego Supercomputer Center and the Protein Data Bank (which he co-directed). At UCSD, Phil was also a Professor of Pharmacology and ultimately an Associate Vice Chancellor.

This tribute, which accompanies an interview in this Special Issue, is not meant to delineate Phil's curriculum vitae or detail his many honors and achievements—e.g., serving as an early president of the *International Society for Computational Biology* and as the first Editor-in-Chief of *PLOS Computational Biology*—but rather to highlight the several ways in which Phil's contributions and leadership in multiple, disparate fields have coalesced as part of a new field of biomolecular data science. For details, note that a brief autobiographical account of Phil is available [2], as are his Wikipedia profile [3], his Ph.D. dissertation [4], and a list of the many scientists [5] whom Phil has trained, mentored and advised over the past four decades (that information is also available as a taxonomic tree [6], fittingly enough). Also, we would be remiss were we not to mention that one can learn what Phil, Monty Python, X-ray crystallography, and the county of Yorkshire, England all have in common by visiting ref [7]. Here, we intentionally intertwine the personal and the professional—as one can gather from even just brief interactions with him, Phil-the-human and Phil-the-scientist are refreshingly one and the same (Figure 1).

Currently a Professor of Biomedical Engineering and the Stephenson Dean of the School of Data Science at UVA, Phil spent much of his career exploring and helping *define* the intersection of biomolecules and computation—as a practicing scientist and as a leader [8] in academia, in open-access academic publishing [9], in the broader open-science movement [10, 11], and in conjunction with government and industry (Phil's role as associate vice chancellor at UCSD concerned "innovation and industrial alliances"). Over the span of Phil's four-decade career, our knowledge of biomolecular structure, dynamics, function and evolution (in both health and disease) has rapidly advanced, often exponentially. *What enabled this?* The staggering advances were enabled, in no small part, by Phil's highly collaborative and foundational work, where three pervasive themes have been: *(i)* a **structural approach** to biological systems, including knowing when to be reductionist and when not to be; *(ii)* the development and application of core **computational methodologies**; and *(iii)* **multidisciplinarity**, to an extreme.

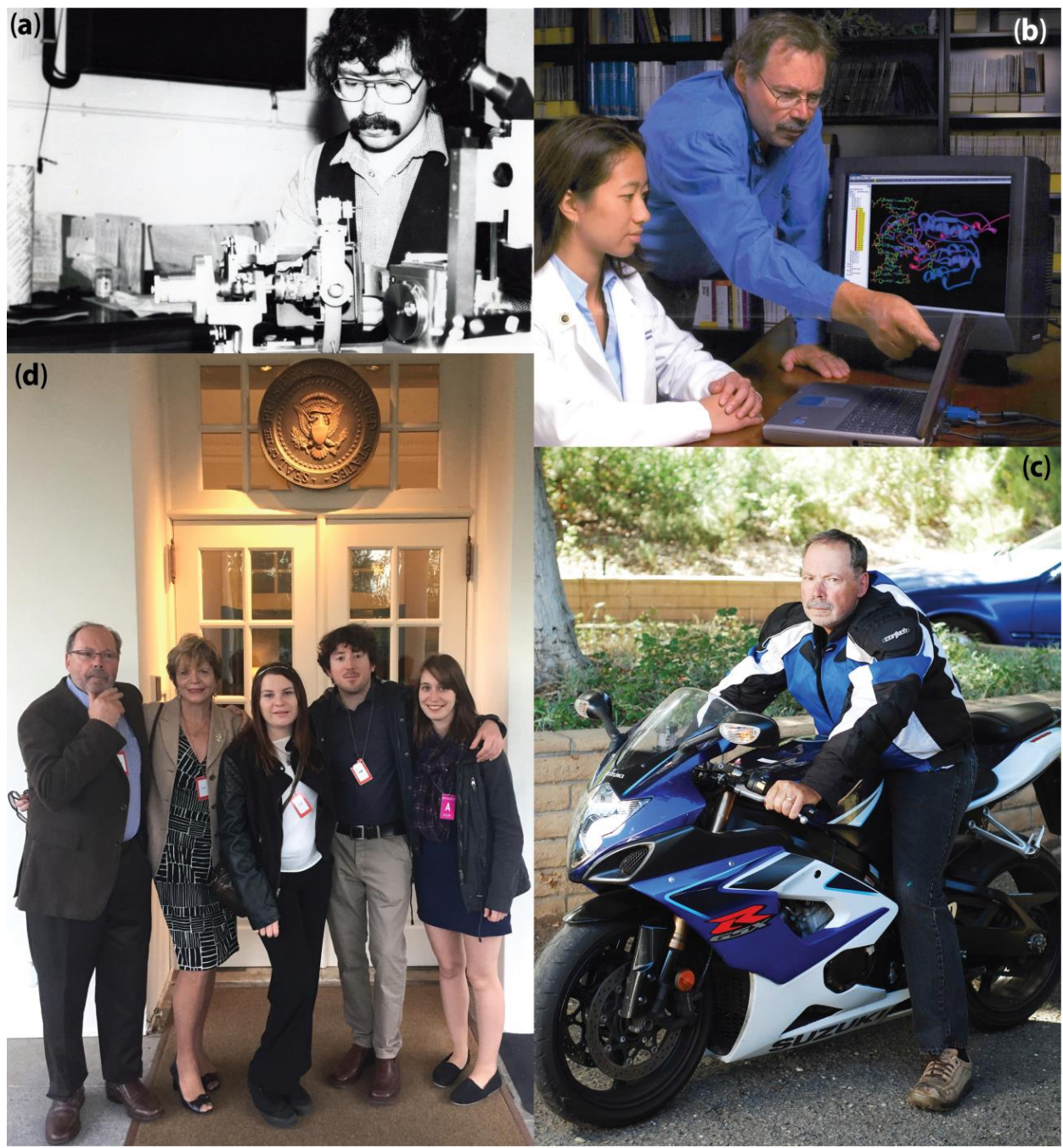

**Figure 1.** Phil's life in science started (a) very hands on, progressed to (b) mentoring, and then he finally (c) took off for the (d) White House with his family. While he's been a scientist for over 40 years, Phil's been an avid cyclist for even longer; at UVA, he's a founding member of the Hells Administrators (https://www.youtube.com/watch?v=ZgtNp1ditzE).

To elaborate the above three points—*biological structure*, *computation*, *multidisciplinarity*—we note that from the start of his career, first in small-molecule crystallography [12] and then in (very) large-molecule structural biology [13], Phil embraced the *key role of three-dimensional structure* [2] as an information-rich bridge between a biomolecule's sequence and its function. (Phil's *Structural Bioinformatics* text, with Jenny Gu, is a mainstay on many researchers' bookshelves [14].) As regards point *(ii)*, a hallmark of Phil's research programs over the years has been the development and application of *computational methodologies & resources*, including state-of-the-art databases (most notably the Protein Data Bank [15]) and associated data standardization, dictionary and exchange approaches, such as the macromolecular crystallographic information file (mmCIF) [16]. Along the way, Phil and his teams created data standards and interoperable tools that were freely disseminated, before this was appreciated and accepted as a scientific best practice, and they developed algorithms and software such as the widely-used combinatorial extension (CE) method for 3D structure alignment [17] and a novel approach to using "sequence order-independent profile–profile alignment" to examine protein functional sites across vast evolutionary distances [18]. Finally, as regards point *(iii)*, computational biology and related areas are well-understood to be *highly* interdisciplinary [19], and here we simply reiterate that Phil was a pioneer in these fields from their inception (before they were 'a thing'). As an extreme example that's specific to Phil, not many scientists have both published research on "ancient shifts in trace metal geochemistry" [20] and written a book on Unix [21]!

In addition to foundational 'basic research' advances, Phil's work and its applications have had significant impact across a vast array of biological and biomedical domains, including early-stage drug discovery [22], molecular evolution [23], immunology [24], and more—resulting in over 350 papers, several books, and nearly 75,000 citations to his work [25]. In recent years, Phil's attention has turned to considering what is possible at the junction of data science and structural biology [26, 27]; notably, Phil's receipt of Microsoft's *Jim Gray Award for eScience* (2010) foretold his move into this area, as that award cited his "*groundbreaking accomplishments in data–intensive science*." All throughout these career milestones, Phil has been unwavering in his support of public service in government and academia, in open scholarship, in research best-practices [28], and in the professional development of all who have crossed his path, from students to peers to colleagues. Indeed, as regards professional development, many readers are likely familiar with the *Ten Simple Rules* (TSR) series that Phil conceived of and initiated 20 years ago. There are now well over 1,000 rules [29], covering everything from strategically forging one's career path in academia, government and industry [30, 31], to winning a Nobel Prize [32], to focused guides on leveraging Git/GitHub [33], to avoiding and resolving conflicts with your colleagues [34]. The full collection of TSRs, which is freely available at ref [35] and organized by topical areas/categories (*Career development*, *Education & mentoring*, etc.), is a testament to how Phil empowers scientists to more effectively navigate the world of very-human scientific activities (papers, talks, careers) that begin where the data-collection and number-crunching end.

Those who have worked with Phil have likely noticed that a pronounced trait in his approach to biosciences, and now data science, is that it is expansive and forward-looking, with a healthy dose of irreverence and provocation [36]—in a word, *visionary*. Phil's interests in recent years have converged upon "biomedical data sciences", which can be viewed as a natural evolution (and synthesis) of bioinformatics, computational biology, structural biology, biophysics, systems biology, and other allied fields [36]. In a real sense, the intense multidisciplinarity of Phil's career foreshadowed a field like biomedical data science. This Special Issue honors Phil by trying to capture his vision as it relates to biomolecules—how this vision arose and what it can encompass, as expressed in a collection of original research papers, perspectives and reviews. We hope that the breadth and depth of the contributions in this Issue convey the spirit of Phil's vision.

Finally, as we honor Phil in this Special Issue, recognizing his role today as Dean of the UVA School of Data Science—the first of its kind in the nation—we close by noting that Phil's vision of biomedical data science can be mapped to four core elements of data science: *Systems*, *Analysis*, *Design* and *Value*. For example, *Systems*, in our context of biomolecular data science, relates to the underlying infrastructure, such as data structures, ontologies, software libraries and tools, that enable discovery. With respect to biomolecules, *Analysis* has been largely dominated by machine learning approaches such as deep learning, for which robust systems and frameworks to access and efficiently utilize training data are critical (e.g., [37]). *Design*, which can refer to human–computer interaction, visualization and so on, has played a vital role throughout the history of structural and computational biology, and now biomolecular data science. Finally, *Value* seeks to optimize the benefit of research for those it serves, from society at large to local communities; here, there exist clear links between drug and therapeutic development, health disparities research, and other realms at the heart of biomolecular and biomedical data sciences.

The papers in this Issue exemplify what a field of biomolecular data sciences can represent, as a fitting tribute to someone who has moved the field forward via his own work and by his steadfast support of many research communities, biomolecular and beyond. In keeping with Phil's mantra, '*Onwards!*'…

**Acknowledgments:** We thank Phil Bourne for providing the photographs used in Figure 1.